\documentclass[aps,twocolumn,superscriptaddress,showpacs]{revtex4}

\usepackage{graphicx}
\usepackage{amsmath}
\usepackage{amssymb}
\usepackage{lscape}
\usepackage{braket}
\usepackage{float,epstopdf}
\usepackage[colorlinks=true, citecolor=red, urlcolor=blue ]{hyperref}
\DeclareMathOperator{\Tr}{Tr}

\begin{document}
\title{Response of entanglement to annealed vis-\`a-vis quenched disorder\\in quantum spin models}
\author{Anindita Bera}
\affiliation{Department of Applied Mathematics, University of Calcutta, 92 A.P.C. Road, Kolkata 700 009, India}
\affiliation{Harish-Chandra Research Institute, HBNI, Chhatnag Road, Jhunsi, Allahabad 211 019, India}
\author{Debasis Sadhukhan}
\affiliation{Harish-Chandra Research Institute, HBNI, Chhatnag Road, Jhunsi, Allahabad 211 019, India}
\author{Debraj Rakshit}
\affiliation{Harish-Chandra Research Institute, HBNI, Chhatnag Road, Jhunsi, Allahabad 211 019, India}
\affiliation{Institute of Physics, Polish Academy of Sciences, Aleja Lotnikow 32/46, PL-02668, Warszawa, Poland}
\author{Aditi Sen(De)}
\affiliation{Harish-Chandra Research Institute, HBNI, Chhatnag Road, Jhunsi, Allahabad 211 019, India}
\author{Ujjwal Sen}
\affiliation{Harish-Chandra Research Institute, HBNI, Chhatnag Road, Jhunsi, Allahabad 211 019, India}
\date{\today}

\begin{abstract}
We investigate bipartite entanglement in random quantum $XY$ models at equilibrium. Depending on the intrinsic time scales associated with  equilibration of the random parameters and measurements associated with observation of the system, we consider two distinct kinds of disorder, namely annealed and quenched disorders. We conduct a comparative study of the effects of disorder on nearest-neighbor entanglement, when the nature of randomness changes from being annealed to quenched. We find that entanglement properties of the annealed and quenched disordered systems are drastically different from each other.
This is realized by identifying the regions of parameter space in which the nearest-neighbor state is entangled, and the regions where a disorder-induced enhancement of entanglement $-$ order-from-disorder $-$ is obtained. We also analyze the response of the quantum phase transition point of the ordered system with the infusion of disorder. 
\end{abstract}
\maketitle

\section{Introduction}
\label{introduction}
For the past few decades, there has been a continued interest to understand effects of randomness, which either appear naturally or incorporated artificially, in many-body systems,  leading to counterintuitive phenomena~\cite{disoredr-review}. Many of the studies are directed towards investigating structural aspects \cite{structural} and cooperative phenomena in random networks, such as disorder induced localization \cite{loc1,loc2,loc3}, high-$T_c$ superconductivity \cite{supercon}, percolation clusters \cite{proclanation}, and rich quantum phase diagrams \cite{phase1,phase2,phase3}. 
Quantum spin models with disordered parameters, in its various incarnations, are known to exhibit many of such phenomena, and can interestingly be implemented in a controlled manner in laboratories dealing with ultracold gases \cite{optical-lattice, Lewenstein_book}. 

In order to obtain a meaningful value of a physical quantity in a disordered system, one must perform a configurational averaging over the disordered parameters. To obtain a general understanding of the response of the different physical quantities due to introduction of disorder, it is useful to consider two contrasting types of disorder, viz.\ annealed and quenched~\cite{disoredr-review, an-qun-0, an-qun-1, an-qun-2, an-qun-3,an-qun-4} which differs by the relative magnitude of two fundamental time-scales in the system.
One of them is the characteristic time, $\tau_m$, during which the system is observed, which includes a possible time-dynamics and a subsequent measurement. The other is the characteristic time, $\tau_c$, required by the disorder in the system to equilibrate. 

Materials, natural or artificial, for which $\tau_m$ and $\tau_c$, corresponding to a certain disorder parameter in that material, are of same or nearby order, are referred to as having annealed disorder. 
To obtain an operationally meaningful observable in this case, we first need to perform a configurational average (over different realizations of the disorder) on the partition function itself and then compute the annealed averaged free energy by considering the logarithm of the averaged partition function, which finally give the annealed averaged physical quantities. 

On the other hand, for  disordered parameters in a certain material, there can be a drastically different  situation, where 
 $\tau_c \gg \tau_m$. 
For this case, any disorder configuration of the system, after being realized, naturally or artificially,  remains effectively frozen throughout the entire observation process.  Hence, the averaging over the random configurations need to be performed \emph{after} the calculation of the physical quantities, for arbitrary given configurations, and the system under consideration is said to posses quenched disorder. 

Computation of quenched averaged physical quantities at equilibrium cannot always be obtained through the computation of the corresponding  quenched averaged free energy, evaluated by taking average over the  logarithm of the partition function for a given random realization. A batch of literature on disordered quantum systems, however, deals with physical quantities that are linear functions of Hermitian operators on the state space, like magnetization and classical correlators. In such cases, an average over the physical quantities, calculated for given disorder configurations, is equivalent to that calculated via a derivative of the disorder-averaged free energy, with the convenient assumption of an interchange between an integral and a derivative. Inequivalence between these two computations creeps in, whenever disorder-averaging is carried out for physical quantities that are not linear functions of Hermitian operators on the state space of a single realization of the given system, an example being entanglement \cite{Horodecki09}. In such cases, 
to obtain the quenched averaged quantity, one needs to compute the physical quantity for each disorder configuration and then average over all the disorder configurations.  

Ideas from quantum information science often shed new light onto collective properties of  many-body systems, having both fundamental and technological implications~\cite{Lewenstein_book, ent-mb}.  Entanglement~\cite{Horodecki09}, in particular, plays a significant role in quantum information processing tasks, prominent examples of which include efficient quantum communication~\cite{dense-code,teleport}, one-way quantum computation \cite{computation}, entanglement-based quantum key distribution \cite{quantum-key},  etc.  Parallely, entanglement has been used for explaining collective quantum phenomena, such as quantum phase transition (QPT) \cite{ent-phase,ent-mb}, superconductivity \cite{ent-supercon}, etc. 

Many of the  works dealing with randomness in quantum networks primarily focus on quenched averaging \cite{Dziarmaga06,wehr,disorder-spin}.  In particular, effects of quenched disorder have recently been studied for quantum information characteristics,  e.g. on  quantum correlations            
\cite{qn-entanglement,Sadhukhan,group}, quantum correlation length \cite{ent-lgt}, monogamy constraints \cite{no-go}, etc. These results crucially depend on the assumption that randomness remains effectively frozen throughout the measurement process. However, given that the time-scale, $\tau_c$, associated with the system, say a spin network, may also be of the order of the measurement time, it is interesting to know how the equilibrium properties of disorder-averaged entanglement are affected due to such a change in the equilibrating dynamics of the disorder in the network. For such cases, as stated before, the entanglement has to be computed via annealed averaging, where the partition function itself has to be averaged over disorder parameters for several random realizations. 
There is only a limited body of  work that attempts to understand the differences in annealed and quenched averaged physical quantities \cite{thorpe, dasgupta, bera2}. In particular, Ref.~\cite{dasgupta} conducts a study on specific heat as a function of temperature in the Ising spin network, in order to find out the changes in equilibrium properties with the change in nature of disorder,  and Ref.~\cite{bera2} investigates spontaneous magnetization in joint presence of both kinds of disorder (annealed and quenched). 
Refs.~\cite{hide1,hide2} carry out studies on annealed averaged entanglement and its witness within a perturbative approach valid for weak disorder strength. It is of interest to understand the general characteristics of annealed averaged  entanglement for arbitrary disorder strength, and at the same time, to build careful understanding about the changes in entanglement properties as the nature of disorder changes from annealed to quenched. Furthermore, it is interesting to compare the effect of these drastically varying models of disorder with ordered systems.

In this work, our prime interest is to carry out a comparative analysis of the response of entanglement properties in many-body systems to the insertion of disorders, which can be annealed or quenched. We choose  the random quantum transverse $XY$ spin chain, where randomness appears either in the interaction part or in the field part. The randomness is drawn from a probability distribution function, which is chosen to be Gaussian in our case. The investigation has been carried out via Jordan-Wigner and Bogoliubov transformation~\cite{lieb61,Barouch70}, which helps in accessing reasonably large systems. We perform a detailed study of bipartite entanglement in the parameter space in presence of disorders, whose strength is varied by tuning the standard deviation so that the distribution of the disorder parameter shifts from narrow to broad. Our analysis shows that the entanglement properties of the system are drastically different depending on the nature of disorders, i.e.\, whether they are annealed or quenched. Interestingly however, the behaviors are qualitatively the same, irrespective of whether the disorder is in the coupling or in the field.   We identify the entangled vs. separable phases for the different systems. We also identify ``enhanced'' phases (in contrast to ``normal'' phases) where two-site nearest neighbor entanglement is enhanced with the introduction of disorder - \`a la ``order-from-disorder''. We find that such phases exist in presence of annealed as well as quenched disordered models. We also analyze the response of the quantum phase transition~\cite{phase2} and factorization points~\cite{factor1,factor2} in the corresponding ordered system with the insertion of disorder.


The rest of the manuscript is structured as follows. Section~\ref{avq} presents the mechanism for obtaining the annealed and quenched averaged values of physical observables. In Sec.~\ref{methodology}, we introduce the models under study, and discuss the methods involved in solving them. Sec.~\ref{discussion} consists of the results for the disorder averaged entanglement. Finally, Sec.~\ref{conclude} provides a conclusion of this work.

\section{Annealed vs. quenched averaging}
\label{avq}
We now briefly discuss about the physical basis as well as the mathematical formalism for obtaining  annealed as well as quenched averaged values of observables. We consider a Hamiltonian $\mathcal{H} = \mathcal{H}\left(\{a_i\}\right)$, where $\{a_i\}$, $i=1,\ldots,N$, are disordered system parameters, so that $\{a_i\}$ can be modeled as independent identically distributed (i.i.d.) random variables following certain probability distributions. The corresponding partition function is given by $\mathcal{Z}\left(\{a_i\}\right)=\Tr\left[ e^{ -\beta \mathcal{H}\left( \{a_i\}\right)  } \right]$, where $\beta=1/(\kappa_B T)$, with $\kappa_B$ and $T$ being the Boltzmann constant and the absolute temperature, respectively. Let us introduce a functional partition function $\tilde{\mathcal{Z}}\left(\{a_i\},\{\lambda_k\}\right)=\Tr\left[ e^{ -\beta \left\{   \mathcal{H}\left( \{a_i\}\right) +\sum_k \lambda_k \mathcal{A}_k \right\} } \right]$, where the additional term $\sum_k \lambda_k \mathcal{A}_k$ represents a ``probe" required for evaluating the expectation values of the physical quantities represented by Hermitian operators. Assuming  $\tilde{\mathcal{Z}}$ to be a sufficiently smooth function of the $\lambda_k$'s, expectation value of the physical quantity of interest $\mathcal{A}_i$ is obtained by differentiating it with respect to $\lambda_i$ at $\lambda_k=0$ $\forall k$. 

As mentioned earlier, annealed disorder corresponds to the situation when the relaxation time for equilibration of disorder is of same or near order of magnitude of the observation time. 
Here the statistical properties of the system at equilibrium is obtained by taking averages of the functional partition function, $\tilde{\mathcal{Z}}$, over several random realizations. 
The functional free energy, $\tilde{\mathcal{F}}^a$,  after performing configurational averaging over the annealed disordered parameters, is given by
\begin{equation}
\tilde{\mathcal{F}}^a (\{\lambda_k\}) = -\frac{1}{\beta} \ln \left\{ \int_{-\infty}^{\infty}   \prod_j {d}a_j \mathcal{P}(a_j) \tilde{\mathcal{Z}}\left(\{a_i\},\{\lambda_k\}\right) \right\}, 
\end{equation}
where $\mathcal{P}(a_j)$ represents the probability distribution function of randomness in the annealed parameter $a_j$. The ``annealed average" in the canonical equilibrium state of a given observable $\mathcal{A}_k$ can finally be obtained as ${\langle \mathcal{A}_k \rangle}^a=\frac{ \partial \tilde{\mathcal{F}}^a}{\partial\lambda_k}\Big{|}_{\{\lambda_i\}=0}$.

Contrary to the annealed disordered parameters, the quenched disordered ones remain effectively frozen throughout the measurement process. In this case, the functional free energy is computed by performing configurational averaging over the quenched disordered parameters of the logarithm of the functional partition function instead of the partition function itself, and reads as
\begin{equation}
\tilde{\mathcal{F}}^q (\{\lambda_k\}) = -\frac{1}{\beta}\int_{-\infty}^{\infty}  \prod_j d{a_j} \mathcal{P}(a_j) \ln \left\{\tilde{\mathcal{Z}}\left(\{a_i\}, \{\lambda_k\}\right) \right\}.
\end{equation}
Suppose now that we wish to compute the ``quenched averaged" value, ${\langle \mathcal{O} \rangle}^q$, of a certain physical characteristic, $\mathcal{O}$, where $\mathcal{O}$ is a non-linear function of physical quantities representable by Hermitian operators $\mathcal{A}_k$ ($\mathcal{O} = \mathcal{O}(\{\mathcal{A}_k\})$). The quenched average ${\langle \mathcal{A}_k \rangle}^q$, of any $\mathcal{A}_k$ can be computed directly, via derivatives of the functional free energy, as ${\langle \mathcal{A}_k \rangle}^q= \frac{\partial \tilde{\mathcal{F}}^q}{\partial \lambda_k}\Big{|}_{\{\lambda_i\}=0}$. However, a physically meaningful quenched averaged value of the observable $\mathcal{O}$ cannot be obtained from the quenched averaged values of the observables $\mathcal{A}_k$. To find ${\langle \mathcal{O}\rangle }^q$, one needs to calculate $\mathcal{A}_k(\{a_i\})$ and the corresponding  ${ \mathcal{O} }(\{{\cal A}_k(\{a_i\})\})$ for each realization of the disorder $\{a_i\}$ and then performs averaging over all such realizations, given by  
\begin{equation}
{\langle \mathcal{O} \rangle}^q=\int_{-\infty}^{\infty}  \prod_j d{a_j} \mathcal{P}(a_j) \mathcal{O}(\{\mathcal{A}_k (\{a_i\})\}).
\end{equation}
On the other hand, in the case of annealed disorder, the averaged out value, ${\langle \mathcal{O} \rangle}^a$, has to be obtained from the annealed averaged values of $\mathcal{A}_k$, i.e., ${\langle \mathcal{O} \rangle}^a =\mathcal{O}(\{{\langle \mathcal{A}_k \rangle}^a\})$.

\section{The model and methodology}
\label{methodology}
In this work, we consider an one-dimensional anisotropic quantum $XY$ model with nearest-neighbor site-dependent interactions in a random transverse magnetic field. The Hamiltonian is given by
\begin{equation}
\label{hami}
\mathcal{H}=\sum_{i=1}^N \frac{\mathcal{J}_i}{4} [(1+\gamma) \sigma_{i}^x \sigma_{i+1}^x+(1-\gamma) \sigma_{i}^y \sigma_{i+1}^y]-\sum_{i=1}^N \frac{h_i}{2} \sigma^z_i,
\end{equation}
where $N$ is the number of lattice sites, $\mathcal{J}_i$ is proportional to the coupling constant between nearest-neighbor sites $i$ and $i+1$, $h_i$ is proportional to the strength of the transverse field at the $i^{\text{th}}$ site, and $\gamma \neq 0$ is the anisotropy parameter. Here $\sigma_i^{\alpha}$ $(\alpha = x, y, z)$  are the Pauli spin matrices at the $i^{\text{th}}$ site. For the homogeneous system, $\mathcal{J}_i$ and $h_i$ are separately equal for all pairs $(i,i+1)$ and for each site, denoted by $\mathcal{J}$ and $h$, respectively. We consider periodic boundary condition, i.e., $\vec{\sigma}_{N+1}=\vec{\sigma_1}$. \\

In the following, we consider  two different cases: 
\vspace{-0.5em}
\begin{enumerate}
\item
The coupling strengths $\mathcal{J}_i$  are randomly chosen.  $\mathcal{J}_i$ are drawn from independently and identically distributed (i.i.d.) Gaussian distributions with mean $\left\langle \mathcal{J} \right\rangle$ and a standard deviation, $\sigma$. However, the system is subjected to an site-independent uniform field, i.e., $h_i = h$ $\forall i$. 
\item
The interaction strength in this case is constant for all pairs, i.e., $\mathcal{J}_i = \mathcal{J}$ $\forall i$. However, the $h_i$ are now  i.i.d. Gaussian random variables with mean $\left\langle {h} \right\rangle$ and the standard deviation, $\sigma$.
\end{enumerate}

%
%
%


The computation of the physical quantities corresponding to the Hamiltonian in Eq.~(\ref{hami}) may need the functional partition function, for which we  introduce a modified Hamiltonian, $\tilde{\mathcal{H}}$, incorporating auxiliary terms (see Sec.~\ref{avq}),  given by
\begin{eqnarray}
\label{ani1}
\tilde{\mathcal{H}} = \frac{1}{4} \sum_{i=1}^N \left[ \gamma_x \tilde{\mathcal{J}}_{i}^x \sigma_{i}^x \sigma_{i+1}^x +
\gamma_y \tilde{\mathcal{J}}_{i}^y \sigma_{i}^y \sigma_{i+1}^y \right] -\frac{1}{2} \sum_{i=1}^N \tilde{h}_i \sigma_i^z,\nonumber\\
\end{eqnarray}
where $\tilde{\mathcal{J}}_{i}^x=(\mathcal{J}_i+4 \lambda^x)$, $\tilde{\mathcal{J}}_{i}^y=(\mathcal{J}_i +4 \lambda^y)$, $\gamma_x=1+\gamma$, $\gamma_y=1-\gamma$, and $ \tilde{h}_i=(h_i+2 \lambda^z)$. Here $\lambda^{\alpha}(\alpha=x,y,z)$ are the coefficients associated with auxiliary terms in $\tilde{\mathcal{H}}$ required for computing observables, such as magnetization and correlators, via derivatives with respect to $\lambda^{\alpha}$ at $\lambda^\alpha=0$ $\forall \alpha$. 
The  $x$-$x$ and $y$-$y$ correlation functions can be obtained via derivatives of the free energy with respect to $\lambda^x$ and $\lambda^y$ respectively, 
while $\lambda^z$ will be used to calculate the magnetization in the $z$-direction by taking derivative of free energy with respect to $\lambda^z$. All the derivatives are to be taken at $\lambda^\alpha=0$ $\forall \alpha$.  

Let us note here that the homogeneous (ordered) quantum $XY$ model can be solved analytically both for finite and infinite chains, and closed forms of magnetization and two-point correlators can also be obtained \cite{Barouch70}. Subsequently, the two-body density matrices can be constructed. Although, the loss of translational symmetry in the disordered $XY$ models restricts the study to finite-sized systems, one can still, in principle, access large finite-sized systems via the Jordan-Wigner transformation \cite{lieb61} which reduces the Hamiltonian in Eq.~(\ref{ani1}) in terms of Fermi operators, $c_i$, as 
\begin{equation}
\label{dekho}
\tilde{\mathcal{H}} = \sum_{i,j=1}^N c_i^\dagger \tilde{A}_{ij} c_j+\frac{1}{2} \sum_{i,j=1}^N (c_i^\dagger \tilde{B}_{ij} c_{j+1}^\dagger+ \mbox{h.c.}).
\end{equation}
Here, $\tilde{A}$ and $\tilde{B}$ are $N \times N$ symmetric and antisymmetric real matrices, respectively, and are given by $\left. \tilde{A}_{ij}=(h_i+2 \lambda^z) \delta_{ij}+(\mathcal{J}_i/2+\lambda^x+\lambda^y) (\delta_{i+1,j}+\delta_{i,j+1})\right.$ and $\left. \tilde{B}_{ij}=({\gamma \mathcal{J}_i}/2+\lambda^x-\lambda^y) (\delta_{i+1,j}-\delta_{i,j+1}) \right.$
with $\tilde{A}_{1N}=\tilde{A}_{N1}={\mathcal{J}_N}/{2}+\lambda^x+\lambda^y$ and $\tilde{B}_{1N}={\gamma \mathcal{J}_N}/{2}+\lambda^x+\lambda^y=-\tilde{B}_{N1}$. 
The Hamiltonian in Eq.~(\ref{dekho}) is further subjected to Fourier and Bogoliubov transformations, given by $\eta_k=\sum_{i=0}^{N-1} (g_{ki} c_i+h_{ki} c_i^\dagger)$,
$\eta_k^\dagger=\sum_{i=0}^{N-1} (g_{ki} c_i^\dagger+h_{ki} c_i)$, after which it reads as 
\begin{equation}
\label{hoteihobe}
\tilde{\mathcal{H}} = \sum_k \Lambda_k \eta_k \eta_k^\dagger+ \mbox{constant},
\end{equation}
where $k=-N/2, N/2+1,\ldots, N/2-1$. Here, $g_{ki}$ and $h_{ki}$ are real numbers, and 
$\eta_k$s obey fermionic anticommutation relations. In this work, we focus on the low temperature properties of the system. 
Note that at absolute zero temperature, the ordered system (i.e.\ ${\cal J}_i={\cal J}$ and $h_i=h~ \forall i)$ undergoes a quantum phase transition at ${\cal J}/{h}=1$, from a paramagnetic (PM) phase for \({\cal J}/{h}<1\) to an antiferromagnetic (AFM) one for \({\cal J}/{h}>1\).


\subsection{Single- and two-site reduced density matrices in presence of annealed disorder}

The computation of entanglement via two-body reduced density matrices requires evaluation of magnetization and two-body correlators. In the following, we discuss how to obtain the physical quantities for the nearest-neighbor sites in presence of annealed disorder. 

In an $N$-site spin system, a $k$-site density matrix is obtained by tracing out all but those $k$ sites. The site-index of these $k$ sites are stored in $\vec{i}=(i_1,i_2,\ldots,i_k)$, with \(i_l\in\{1,2,\ldots,N\}\), \(l=1,2,\ldots,k\).
The general form of the $k$-site reduced density matrix can be written as 
\begin{align}
\label{red-den-gen-form}
&\rho_{\vec{i}}=\frac{\mathbb{I}_{2^k}}{2^k} + 
&\sum_{i=1}^{k}\frac{1}{2^i} \sum_{\vec{\alpha} \ne 0 } C_{\vec{\alpha}} ~ \sigma^{\alpha_{i_1}}_{i_1} \otimes \sigma^{\alpha_{i_2}}_{i_2} \ldots \otimes \sigma^{\alpha_{i_k}}_{i_k}.
\end{align}
Here $\vec{\alpha}=(\alpha_{i_1}, \ldots, \alpha_{i_k})$, where each entry in $\vec{\alpha}$ can be any of $\{0,x,y,z\}$, and we set $\sigma^0=\mathbb{I}_2$. Note that 
$ C_{\vec{\alpha}}=\Tr{ (\sigma^{\alpha_{i_1}}_{i_1} \otimes  \ldots \otimes \sigma^{\alpha_{i_k}}_{i_k} \rho_{\vec{i}})}$. If there are $p$  non-zero entries in $\vec{\alpha}$, then $ C_{\vec{\alpha}}$ represents a $p$-body correlator. In particular, when there is a single non-zero entry in $\vec{\alpha}$, then $C_{0,\ldots,\alpha_{i_l},\ldots,0} = m^{i_l}_{\alpha_{i_l}}$ ($\alpha_{i_l}\ne 0$) represents the magnetization at site $i_l$.
Let us assume that the $N$-body density matrix, $\rho_{N}$, has global phase flip symmetry $\left(\left[\rho_{N}, ({\sigma^z})^{\otimes{N}}\right]=0\right)$ implying $\rho_{N} =  (\sigma^z)^{\otimes{N}}  \rho_N (\sigma^z)^{\otimes{N}}$. This, e.g., is the case for the ground state of $\mathcal{H}$. Consequently, since trace, including partial trace, is basis independent, the $k$-site reduced density matrix is given by 
\begin{eqnarray}
\label{red_syym}
\rho_{\vec{i}}&=&\Tr_{\bar{\vec{i}}} \rho_N
=\sum_{\bar{k}} \langle \bar{k}| (\sigma^z)^{\otimes{N}}  \rho_N  (\sigma^z)^{\otimes{N}} |\bar{k} \rangle \nonumber\\
&=&(\sigma^z)^{\otimes k} \left( \sum_{\bar{k}} \langle \bar{k}|(\sigma^z)^{\otimes \bar{k}} \rho_N   (\sigma^z)^{\otimes \bar{k}} |\bar{k} \rangle \right) (\sigma^z)^{\otimes k} \nonumber\\
&=&(\sigma^z)^{\otimes k} \rho_{\vec{i}} (\sigma^z)^{\otimes k},
\end{eqnarray} \\
where $\bar{\vec{i}}$ denotes the nodes which are traced out.


We now assume that the annealed averaged state is an analytic function of the set of Hamiltonians for different realizations of the annealed disordered parameters in $\mathcal{H}$. The global phase flip symmetry holds for the zero-temperature state as for the canonical equilibrium state corresponding to $\tilde{\mathcal{H}}$ for every particular realization  of disorder and hence the symmetry also holds for the annealed averaged state, $\rho^a$. For single site, Eq.~(\ref{red_syym}) reduces to $\left[\rho_{1}^a, \sigma^z\right]=0$. Using the general form of the single-site density matrix $\rho_{1}^a$, (see Eq.~(\ref{red-den-gen-form})), one obtains
\begin{equation}
\label{science}
m_x^a \sigma^x+m_y^a \sigma^y=0,
\end{equation}
which implies $m_x^a=m_y^a=0$, as $\sigma^x$ and $\sigma^y$ are linearly independent generators. Here, we have used the notation $m_\alpha^a=\langle \sigma_1^\alpha \rangle^a$. Similarly, for the two-site reduced density matrix, $\rho_{12}^a$ of the annealed averaged state, $\left[\rho_{12}^a, (\sigma^z)^{\otimes 2} \right]=0$, which in turn implies $C_{xz}^a=C_{zx}^a=C_{yz}^a=C_{zy}^a=0$. Again, we have used the notation $C_{\alpha_1 \alpha_2}^a=\langle \sigma_1^{\alpha_1} \otimes \sigma_2^{\alpha_2} \rangle^a$. 
We now go over to the consideration of the \(xy\) and \(yx\) correlators.
%
%
The annealed averaged correlation, $C_{xy}^a$, is given by
 $$C_{xy}^a =\lim_{c_{xy}\to 0} \frac{\partial}{\partial c_{xy}} \ln \int \mathcal{P}(\mathcal{J}) d\mathcal{J}  \tilde{\mathcal{Z'}},$$ where $ \tilde{\mathcal{Z'}}$ is the partition function corresponding to the Hamiltonian  $\tilde{\mathcal{H'}}= \tilde{\mathcal{H}}(\mathcal{J})+c_{xy} \sigma^x \otimes \sigma^y$, where $\tilde{\mathcal{H}}$ is given in Eq.~(\ref{ani1}). $\mathcal{J}$ denotes the aggregate of all the annealed disordered parameters in the Hamiltonian. 
 For the canonical equilibrium state of the \emph{ordered} system, this approach implies the corresponding \(C_{xy} = 0\), by using 
 the fact that the density matrix is an analytic function of the corresponding Hamiltonian. The same 
 does not seem to go through for the annealed disordered case, even if we assume that the annealed state is an analytic function of the aggregate of annealed disordered Hamiltonians. We therefore resort to numerical simulations, and indeed we find by exact diagonalizations of up to 
 11 quantum spin-1/2 systems that \(C_{xy}^a = 0\). The case is similar for \(C_{yx}^a\). On the basis of these evidences, we assume that \(C_{xy}^a = C_{yx}^a=0\) for larger systems as well.
 %
 %
Hence $m_z^a$, $C_{xx}^a$, $C_{yy}^a$, and $C_{zz}^a$ emerge as the non-zero quantities for the annealed averaged state.
Additionally, let us show that 
$C_{zz}^a=\Tr(\rho_{12}^a \sigma_1^z \otimes \sigma^z_2)$ can be expressed in terms of the operators, $A_i$ and $B_i$, as $C_{zz}=\Tr (\rho_{12}^a A_1 B_1 A_2 B_2)$ and hence can be written in terms of $m_z^a, C_{xx}^a$ and $C_{yy}^a$. A further decomposition by means of Wick's theorem leads to 
\begin{eqnarray}
C_{zz}^a &=& \Tr (\rho_{12}^a A_1 B_1) \Tr (\rho_{12}^a A_2 B_2) \nonumber\\
&-&\Tr(\rho_{12}^a A_1 A_2) \Tr (\rho_{12}^a B_1 B_2)\nonumber\\
&-& \Tr(\rho_{12}^a B_2 A_1) \Tr(\rho_{12}^a B_1 A_2), \nonumber\\
&=& (m_z^a)_1 (m^a_z)_2-C^a_{yy} C^a_{xx},
\end{eqnarray}
where $(m_z^a)_1$ and $(m_z^a)_2$ are the annealed averaged magnetizations of sites 1 and 2 respectively.

For the quenched system, the disordered Hamiltonian again satisfies the global phase flip symmetry, and so a relation of the form (\ref{science}) can be obtained for each realization of the quenched disordered Hamiltonian.  This immediately provides \(m_x^q=m_y^q=0\). Global phase flip symmetry applied to the two-site state implies 
$C_{xz}^q=C_{zx}^q=C_{yz}^q=C_{zy}^q=0$. Since the Hamiltonian, for every realization of the quenched disordered parameters, has real entries when written in the computational basis, 
the \(xy\)- and \(yx\)-correlators vanish in the canonical equilibrium state, for that disorder realization. This of course implies that the quenched averages, \(C_{xy}^q\) and \(C_{yx}^q\) are also vanishing. So, for the quenched system, we again find that the two-site state possesses only 
the z-magnetization and the diagonal correlators. 
A similar argument holds for the homogeneous system as well.

\section{Annealed vs. quenched entanglement in random $XY$ model}
\label{discussion}
In a multiparty system, bipartite entanglement between two subsystems characterizes collective properties of the many-body system, and at the same time, its existence may ensure implementation of efficient quantum information processing tasks. In this work, we choose concurrence~\cite{hill} as the measure of bipartite (two-qubit) entanglement for analyzing the behavior of nearest-neighbor states in annealed as well as quenched disordered systems, for identifying their similarities and differences, affecting their relative utilities. We also compare the physical properties of the disordered systems with the same in the corresponding ordered systems.

For an arbitrary two-qubit density matrix, $\rho$, concurrence is defined as $C(\rho) = \max\{0, \lambda_1 - \lambda_2 - \lambda_3 - \lambda_4\}$, where $\lambda_i$'s are eigenvalues of the Hermitian matrix $R=\sqrt{\sqrt{\rho} \tilde{\rho} \sqrt{\rho}}$, in descending order. Here $\tilde{\rho}=(\sigma_y \otimes \sigma_y) \rho^* (\sigma_y \otimes \sigma_y)$, where $\rho^*$ is the complex conjugate of $\rho$ in the computational basis. 


The means of the distributions of the random parameters in the disordered systems are adjusted to be identical to the corresponding parameters of the homogeneous system. Moreover, the standard deviations corresponding to different types of disorder for the same physical parameter (e.g., the coupling strength) in different systems are taken to be equal, so that the effects due to the disorders can be compared effectively. Below, for a given observable $\mathcal{O}$, we compare between 
$\langle \mathcal{O} \rangle^a$, $\langle \mathcal{O} \rangle^q$, and $\langle \mathcal{O} \rangle$, which, respectively, are averages in the equilibrium state  for annealed disorder in the coupling, quenched disorder in the same, and a constant (site-independent) coupling. 

We now investigate the effects of annealed and quenched disorders on entanglement in the transverse field quantum $XY$ spin model. Let us first consider the case where randomness is present only in the interaction term, while $h_i=h~\forall i$. 
 Our prime interest is to study the changes in the entanglement properties of the system, governed by the Hamiltonian given in Eq.~(\ref{hami}), as the nature of disorder changes from annealed to quenched. The random interactions, $\mathcal{J}_i$, are chosen to be independently and identically distributed Gaussian probability distributions with mean $\langle \mathcal{J} \rangle$ and standard deviation $\tilde{\sigma}$, where  $\langle \mathcal{J} \rangle$ and $\tilde{\sigma}$ are site-independent. Note that $\tilde{\sigma}$ represents the disorder strength. 
The associated probability density function, $\mathcal{P}(\mathcal{J}_i)$, of the disordered parameters, $\mathcal{J}_i$, is given by $\mathcal{P}(\mathcal{J}_i)=\frac{1}{\tilde{\sigma}\sqrt{2 \pi }}\exp[\frac{-(\mathcal{J}_i-\langle \mathcal{J} \rangle)^2}{2 \tilde{\sigma}^2}]$. 
In order to keep the disorder averaged quantities and corresponding ordered ones on the same footing, the concurrence in the homogeneous system is calculated between the same sites by setting  $\mathcal{J}_i=\langle \mathcal{J} \rangle~\forall i$.
To study any disorder averaged physical quantity, one typically requires a few thousand random realizations in order to obtain converged values via configurational averaging. Throughout this work, the averaged out quantities for the disordered systems are calculated by performing averaging over $10^4$ random realizations.

\begin{figure}[t]
\includegraphics[angle=0,width=0.48\textwidth]{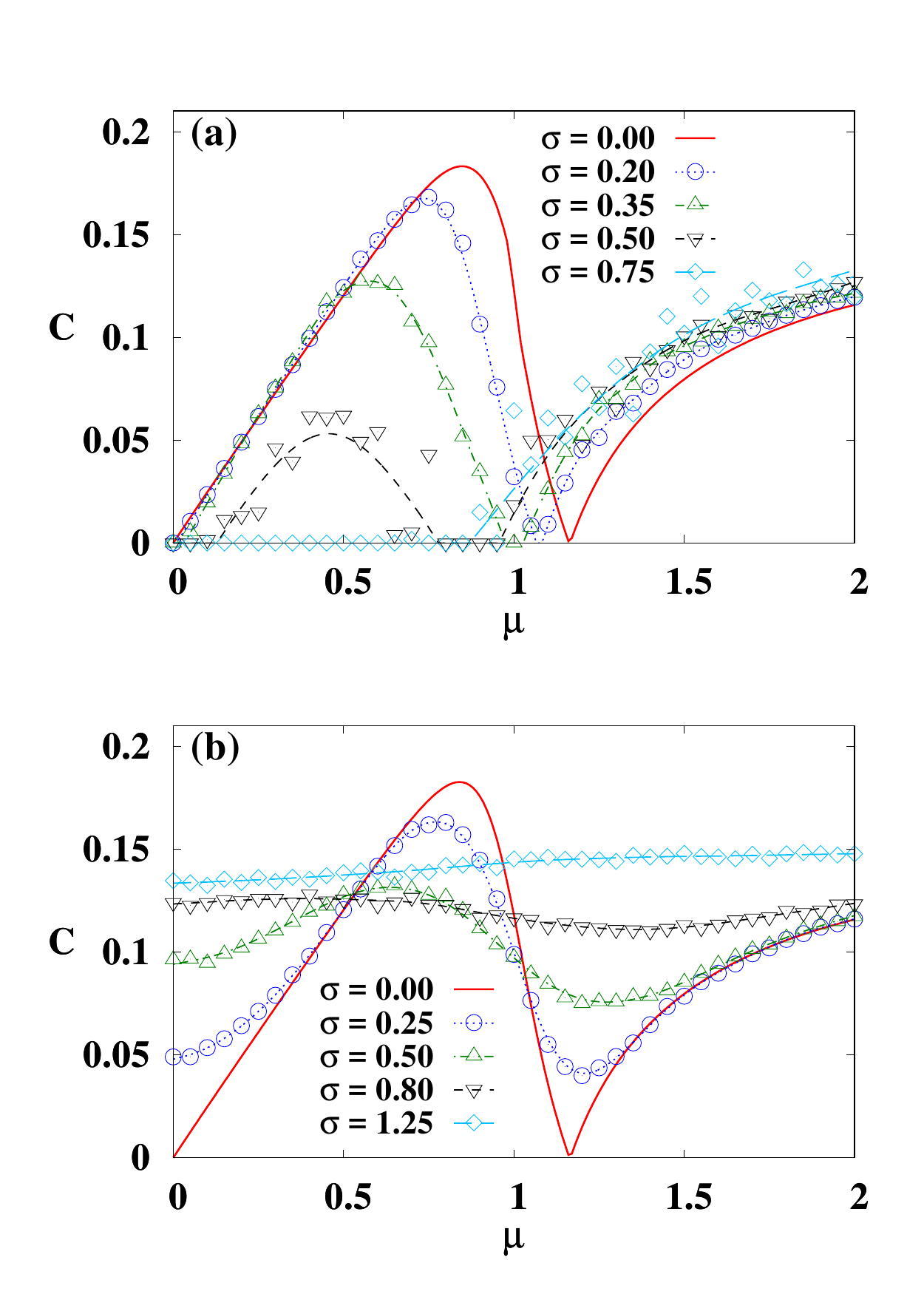}
\caption{(color online) Response of entanglement to annealed and quenched dirorders, respectively. We use concurrence as the measure of entanglement, and plot the concurrence, $C$, as a function of $\mu~(=\frac{\langle \mathcal{J} \rangle}{h})$ for different disorder strengths, $\sigma$, in presence of (a) annealed and (b) quenched disorder in the coupling. In both the panels, the red solid line corresponds to the homogeneous case, i.e.  $\sigma = 0,$ while the other curves are for different disorder strengths as indicated in the legends. For the annealed disorder,  fitting curves are obtained using annealed averaged concurrence data, which in turn is obtained from spline interpolation of annealed averaged magnetization and classical correlations, while for the quenched disorder, the fitted curves are the spline interpolation of the quenched averaged concurrence. The quenched averaged concurrence is obtained by calculating it for a given disorder realization, and then pertaining the average.  
Averaging is performed over $10^4$ random realizations, and we have checked that all the physical quantities considered here, have converged much before that sample size.  The vertical axes are measured in ebits, while the horizontal axes are dimensionless. We choose $N=50$ and $\gamma=0.5$ for all figures in this paper, and $\beta h =20$ for Figs.~\ref{fig1}-\ref{fig_OfD_phase}.}
\label{fig1}
\end{figure}

\subsection{Case 1:~ Transverse $XY$ model with random interaction}

Fig.~\ref{fig1} shows the behavior of concurrence, $C$, as a function of $\mu = \langle {\cal J}\rangle/h$ in the $XY$ spin chain with $\gamma = 0.5$ in presence of annealed (Fig.~\ref{fig1}(a)) and quenched disorders (Fig.~\ref{fig1}(b)), respectively. 
 On the other hand, for the homogeneous system, $\mu$ is site-independent and it is 
 set at $\langle \mathcal{J} \rangle/h$. Let us start with the homogeneous system.
 The red solid lines in Figs.~\ref{fig1}(a) and \ref{fig1}(b) show the trends of concurrence for the homogeneous $XY$ spin chain for the canonical equilibrium state. 
The states are entangled except at the $ \mu = 0$ point and the factorization point \cite{factor1,factor2}, with the latter being given by $\mu=\mu_f \equiv 1/\sqrt{1-\gamma^2}$.

In case of an annealed disordered system, to obtain the trends of concurrence, 
we first find the annealed averaged (transverse) magnetization and classical correlators, and use them to 
construct the annealed averaged two-site density matrix. The annealed averaged entanglement is 
then obtained from this annealed averaged density matrix. 
It is important to stress a technical point about obtaining the annealed averaged magnetization and 
classical correlators. For obtaining the behavior of, say, the annealed averaged magnetization as a 
function of \(\mu\),  
we begin by calculating the values of annealed averaged magnetization for a certain chosen set of 
points in the interval of interest on the 
\(\mu\)-axis. (The other parameters of the parameter space, given by $\big(\gamma, \mu, \tilde{\sigma}, \beta\big)$, are suitably chosen.) These values are then fitted to a profile using spline interpolation techniques. The same is 
done for all the classical correlators. The annealed entanglements for the above chosen set of points on the \(\mu\)-axis are then obtained by using the spline-interpolated annealed-averaged 
magnetizations and classical correlators.

We find that the introduction of annealed disorder affects bipartite entanglement significantly, as can be seen in Fig.~\ref{fig1}(a). While in some region, entanglement increases in the presence of annealed disorder, it can get suppressed  due to disorder in other regions. 
%
Below, we divide the axis of rescaled disorder strength, \(\sigma = \tilde{\sigma}/h\), into 
two regions depending on whether the annealed averaged entanglement revives once or twice as we move along the \(\mu\)-axis. 
Here we fix $\gamma=0.5$. A change of $\gamma$ can quantitatively change the boundaries of these regions, although the qualitative behavior in the $(\mu,\sigma)$ plane remains the same for $\gamma\ne 0$.

{\it Region 1} ($\sigma<\sigma_c$) -- 
This region is defined as that in which there are two ``revivals''  of entanglement 
after its ``collapse'' to zero, as we move along the \(\mu\)-axis, for a given value of 
\(\sigma\). 
A ``revival'' is defined as the appearance of (non-zero) entanglement after it becomes zero (``collapse'').
The maximum value of \(\sigma\) for which two revivals occur is denoted by \(\sigma_c\).
The first revival of entanglement on the 
\(\mu\)-axis always leads to a shrinking of entanglement with increasing \(\sigma\).
The situation for the second revival is richer, and can lead to both shrinking as well as 
enhancement of entanglement with \(\sigma\). 
The end of the first revival and the beginning of the second happens in the vicinity of the 
quantum phase transition point and the factorization point of the corresponding ordered system. 
For the system under consideration, we find $\sigma_c \approx 0.6$.
%
Note that the defining property of the critical disorder strength can equivalently be taken as the number of entangled segments on any 
line of constant \(\sigma\).


{\it Region 2} ($\sigma>\sigma_c)$ --  
This region is the one that is complementary to Region 1, and there is a single revival of entanglement when we walk along the \(\mu\)-axis. Here we obtain both shrinking as well as enhancement of entanglement with increasing \(\sigma\).
 For $\sigma \lesssim 0.8$, we find that the point of entanglement revival shifts towards the low
 \(\mu\) region. The opposite happens for higher \(\sigma\).
Note that for a high $\sigma$, a finite amount of entanglement survives in presence of the annealed disorder only if the system is deep in the AFM phase of the corresponding ordered system.

\begin{figure}[t]
\includegraphics[angle=0,width=0.485\textwidth]{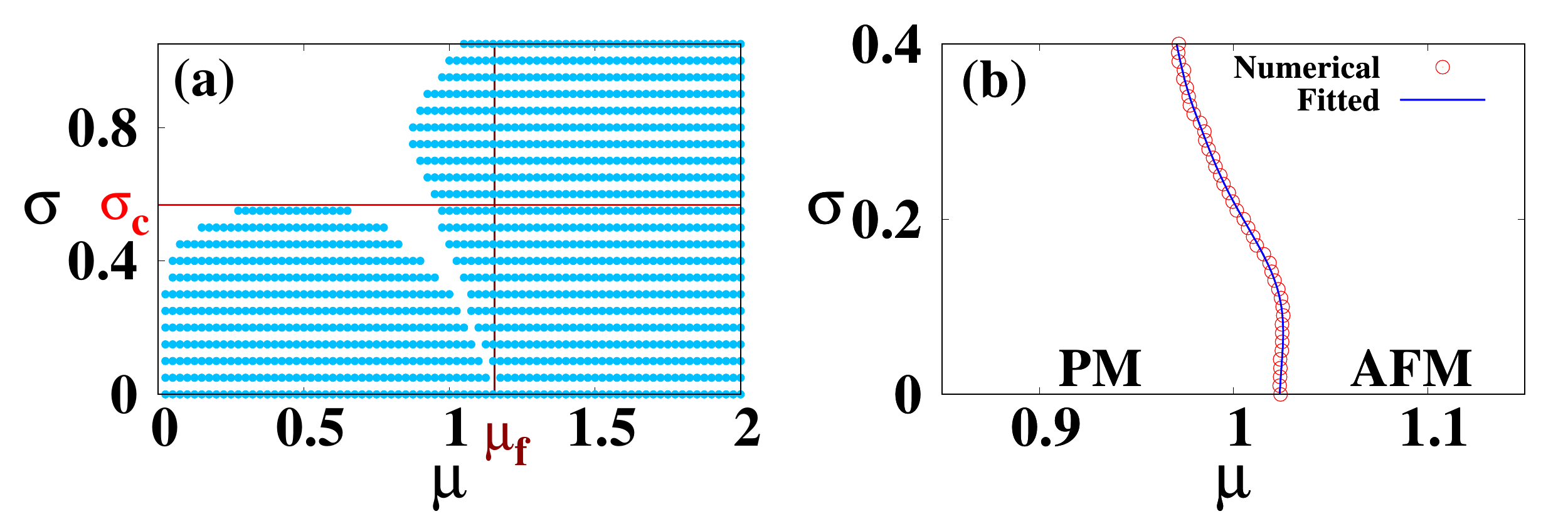}
\caption{ (color online) An illustration of (a) entangled and separable phases in annealed disordered system, and (b) shifting of AFM-PM transition points in quenched disorder system in the parameter space of $(\mu,\sigma)$. The disorders are in the couplings.  In panel (a), the white region represents the separable phase of the annealed disordered system. The separable phase identifies the parametric regime with vanishing annealed averaged entanglement. In panel (b), we consider quenched disorder. For $\sigma=0$, the system corresponds to the homogeneous model. The AFM-PM transition point at $\mu = 1$ for $\beta=\infty$ shifts to $\mu = 1.0243$ for $\beta h= 20$ and $N=50$. With the introduction of quenched disorder the ``transition'' point (see text) drifts from its $\sigma=0$ value.  
All quantities on the axes are dimensionless.
}
\label{fig_phase1}
\end{figure}

The above observation is summarized in Fig.~\ref{fig_phase1}(a), where the tract with blue stripes and that which is white, respectively represent the entangled and separable phases for the annealed disordered systems in the $(\mu$--$\sigma)$ plane. 
There are two points on the \(\sigma=0\) line (ordered system), namely, \(\mu=0\) and \(\mu=\mu_f\), for which the entanglement 
vanishes in the zero temperature state. Insertion of annealed disorder indicates that we are pushing away from the \(\sigma=0\) line on the \((\sigma\)--\(\mu)\) plane, and the separability-entanglement features in the system responds by creating two ``rivers'' of separable states. 
More precisely, the two zero-entanglement points on the \(\sigma=0\) line develops into 
two finite intervals of zero entanglement on any line of (non-zero) constant \(\sigma\), as long as the constant is less than \(\sigma_c\).   For \(\sigma>\sigma_c\), the two rivers meet to create a 
``separable sea''. 

Let us now move to the case when we consider a quenched disordered system.
The bipartite entanglement in this case turns out to be drastically different from that in the annealed case. 
Unlike annealed averaged entanglement, the quenched averaged entanglement between nearest-neighbor sites remains non-zero throughout the entire parametric stretch of $\mu$, irrespective of the strength of disorder, however small (but non-zero). See Fig.~\ref{fig1}(b). In particular, we observe that with increase of the rescaled quenched disorder strength, $\sigma$, entanglement is generated even at the points having vanishing entanglement in the corresponding ordered Hamiltonian.
As we increase $\sigma$, there is a change in the pattern of entanglement with $\mu$ -- it gets flattened as a function of $\mu$. For a high enough $\sigma$ (in our case, $\sigma \gtrsim 1.3$ ), it  saturates to a moderate non-zero value for the entire parametric stretch of $\mu$. This ``frozen'' entanglement   with respect to $\mu$ for high enough $\sigma$ is expected, because with increasing $\sigma$, the stretch of the Gaussian distribution of the disordered parameter increases and finally  the random configurations, $\mathcal{H}(\{{\cal J}_i/h\})$, are effectively distributed over the entire parametric regime of $\mu$, and thus becomes independent of the mean of the distribution.

\begin{figure}[t]
\includegraphics[angle=0,width=0.48\textwidth]{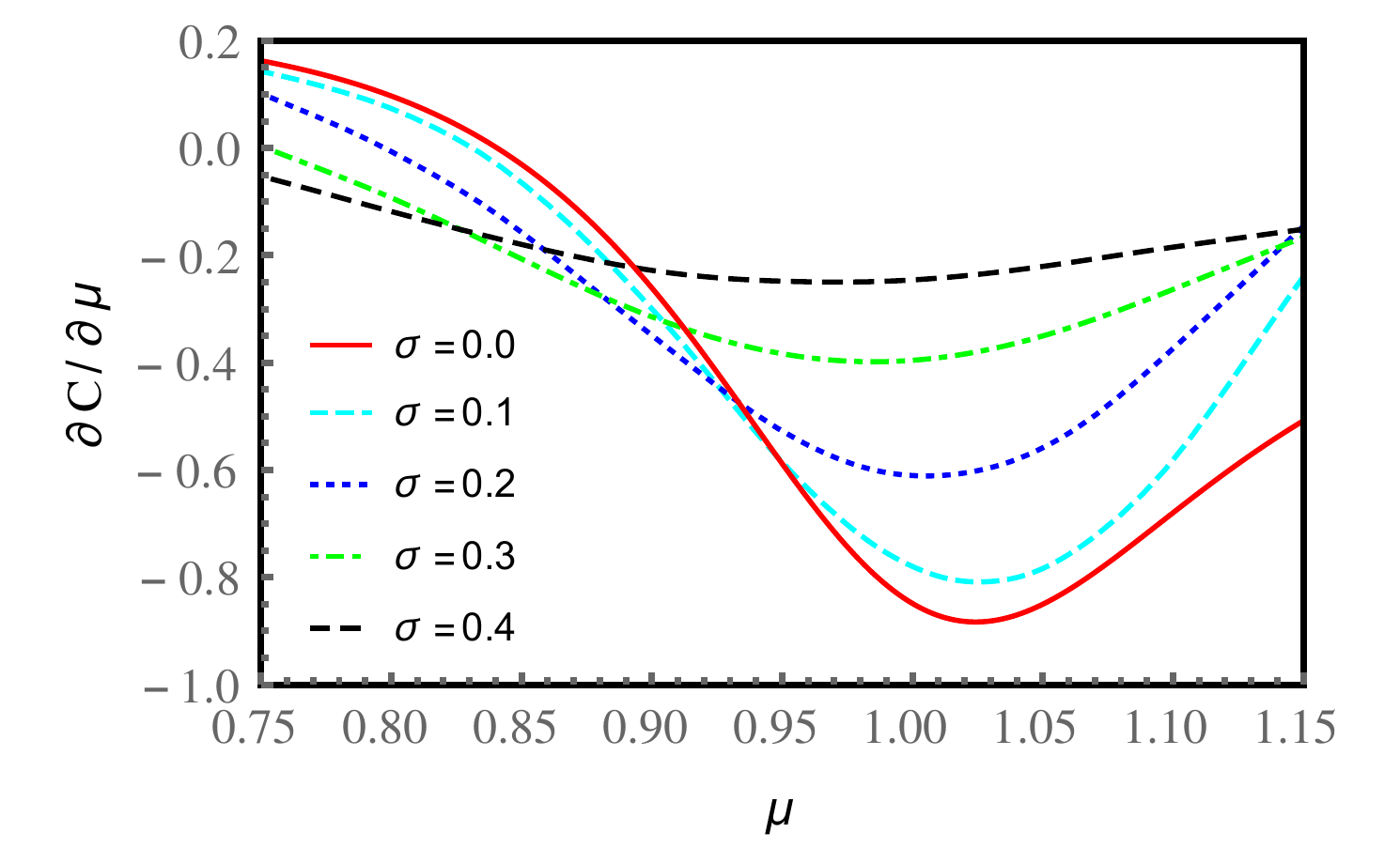}
\caption{(color online) Response of disorder in coupling on a quantum phase transition. The derivative of concurrence, $\frac{\partial C}{\partial \mu}$, is plotted as a function of $\mu~(=\langle \mathcal{J}\rangle /h)$ in presence of quenched disorder in the coupling terms, for different disorder strengths, $\sigma$. The (red) solid line again corresponds to the homogeneous case i.e. with $\sigma = 0,$ while the other curves are for different disorder strength presented in the legends. For a better illustration, only the fitted curves are plotted. The value of $\mu$ at the minimum of a certain curve is interpreted as the  post-response quantum phase transition point for the corresponding $\sigma$. Concurrence is measured in ebits, while $\mu$ is dimensionless. 
}
\label{fig_derivative}
\end{figure}

Bipartite entanglement of the zero-temperature state can characterize the transition at $\mu = 1$ by showing a kink in its derivative. In our case, to maintain consistency with the annealed case, we choose the relative inverse temperature at $\beta h=20$, which  mimics the zero-temperature characteristic of the system. However, we find that due to finite temperature ($\beta h=20$) and finite system size ($N=50$), the kink in the derivative of nearest-neighbor entanglement in the zero-temperature state
of the ordered system shifts to $\mu = 1.0243$ (correct to four decimal places). In Figs. \ref{fig1}, \ref{fig_derivative} and \ref{fig_phase1h}, the red solid line corresponds to concurrence or the derivative of concurrence in the homogeneous case. 
Since we are dealing with finite temperature and finite system size, instead of a sharp kink, we observe a ``blunt'' minimum,  characterizing the QPT, in Fig.~\ref{fig_derivative}. The transition point at $\mu = 1.0243$ can be found from the minimum point in the derivative of concurrence (red solid line of Fig. \ref{fig_derivative}). 

Moreover, we find that the QPT, present in the corresponding ordered system, shifts, as the minimum of $\frac{dC}{d\mu}$ indicates. Specifically,
for small values of quenched disorder strength ($\sigma \lesssim 0.1$), the minimum of $\frac{dC}{d\mu}$ at $\mu=\mu_{\min}$  
remains almost a constant as a function of \(\sigma\),
but afterwards (i.e., for higher \(\sigma\)), it shifts towards the left (PM phase of the ordered system). See Fig.~\ref{fig_phase1}(b). For a relatively stronger disorder ($\sigma > 0.4$), the curves for the quenched disordered concurrence gets flattened and a prominent minimum is thus unavailable.

\begin{figure}[t]
\includegraphics[angle=0,width=0.485\textwidth]{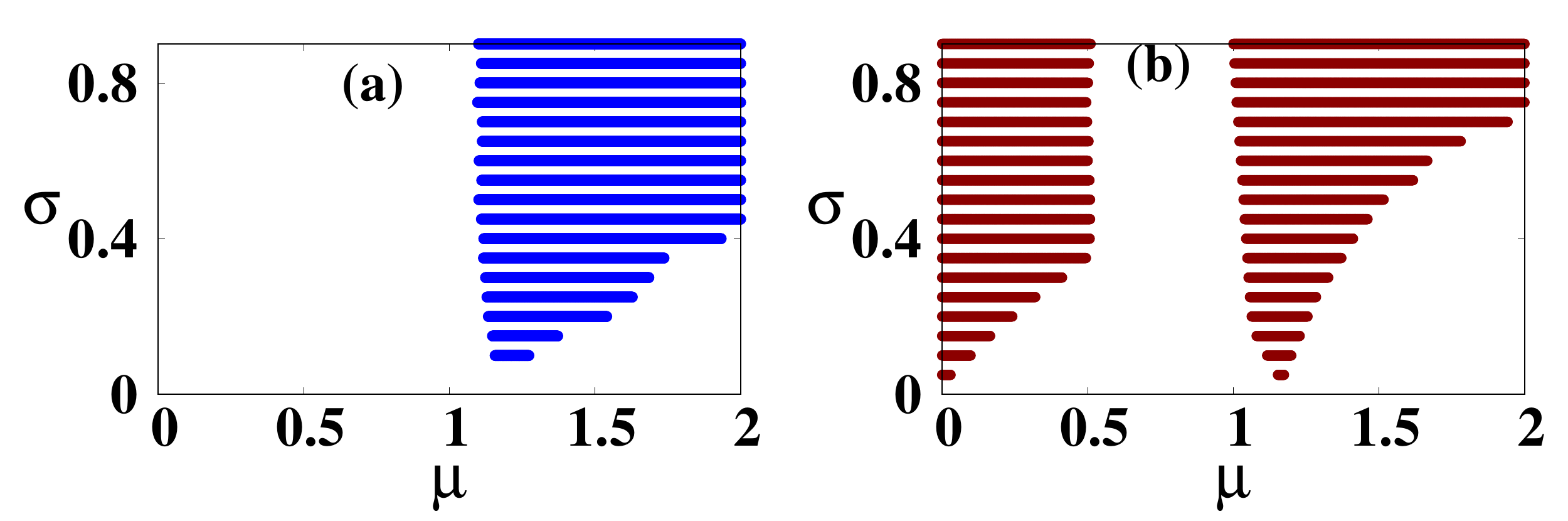}
\caption{(color online) Phases with order-from-disorder separated from those without so. We present  illustrations of enhanced and normal phases in (a) annealed and (b) quenched disordered systems in the parameter space of $(\mu,\sigma)$. The disorders are in the couplings. In both panels, the white  region represents the {\it normal} phase, where no order-from-disorder is present. In these regions, the value of the annealed or quenched averaged entanglement is lower than the same in  the corresponding homogeneous system. The (colored) striped regions, named as {\it enhanced} phases,  correspond to the parametric regimes with order-from-disorder phenomenon for entanglement. 
The disorder averages are checked for convergence up to the third decimal place, and therefore the phase changes are decided with a tolerance of $10^{-3}$. All axes in the plot are dimensionless. Entanglement is measured in ebits.
}
\label{fig_OfD_phase}
\end{figure}

\subsubsection{Disorder-induced enhancement of entanglement}

Presence of any weak quenched disorder makes the entire system entangled even when in the   neighborhood of $\mu =0$, 
implying that disorder helps the system to possess a higher amount of entanglement in  comparison to the corresponding ordered one.  This feature of ``disorder-induced-enhancement" is also known as  ``order-from-disorder" phenomenon, and has been elaborately studied in context of several physical quantities in the past \cite{wehr,Sadhukhan,ent-lgt,od-dis,bera}, especially in quenched disordered systems.  In fact, there are large parameter stretches over which such phenomenon occurs. We define an ``enhanced phase" as one which supports the disorder-induced-enhancement phenomenon. In contrast, ``normal phase" is marked by deterioration of entanglement in presence of randomness. 
These two phases in the annealed as well as quenched disordered systems are presented in Fig.~\ref{fig_OfD_phase}.
For the annealed disorder, the enhanced region appears only in the AFM phase of the ordered system (the blue striped region of Fig. \ref{fig_OfD_phase}(a)), while for quenched disorder, the same occurs in both  AFM and PM phases (the maroon striped regions of Fig. \ref{fig_OfD_phase}(b)).


 In case of the annealed disordered system, the enhanced phase is inside the half-plane $\mu>1$. This corresponds to 
 the AFM phase in the ordered system. Walking along a constant \(\sigma\) line, 
 from low to high values of \(\mu\),
 one encounters the enhanced phase at a value of \(\mu\) that is greater than unity, and that is almost independent of \(\sigma\). 
 Re-entry into the normal phase, however, depends on \(\sigma\). See Fig. \ref{fig_OfD_phase}(a)).


In the quenched disorder scenario, enhanced phases appear in both AFM and PM phases of the corresponding ordered system. 
We find that for a given  $\sigma$, the system is in the enhanced phase for $0<\mu<\mu^\sigma_1$ and \(\mu^\sigma_2 < \mu < \mu^\sigma_3\) with 
\(\mu^\sigma_1 < \mu^\sigma_2 < \mu^\sigma_3\).
Interestingly, for high enough $\sigma$, $\mu^\sigma_1$ and \(\mu^\sigma_2\) are almost independent of $\sigma$. 
As depicted in Fig.~\ref{fig_OfD_phase}(b), the length of the enhanced phase on a constant \(\sigma\) line, increases with $\sigma$. 
Our analysis shows that in the enhanced phase, the two-site entanglement  increases in magnitude with increasing disorder strength before attaining a  saturating value.


It is worth mentioning here that although the whole analysis is carried out for $N=50$, we have checked that the behaviors of entanglement remain unaltered for larger system sizes. 
Moreover, although the illustrations have been presented for a specific value of the anisotropic constant, viz. $\gamma = 0.5$, for which the factorization point is at $\mu = \mu_f = 1.1547$, similar analysis performed for other values of $\gamma$ confirms that the qualitative features remain unaltered irrespective of the value  of the anisotropy.

\subsection{Case 2:~$XY$ model in random transverse field}

\begin{figure}[t]
\includegraphics[angle=0,width=0.5\textwidth]{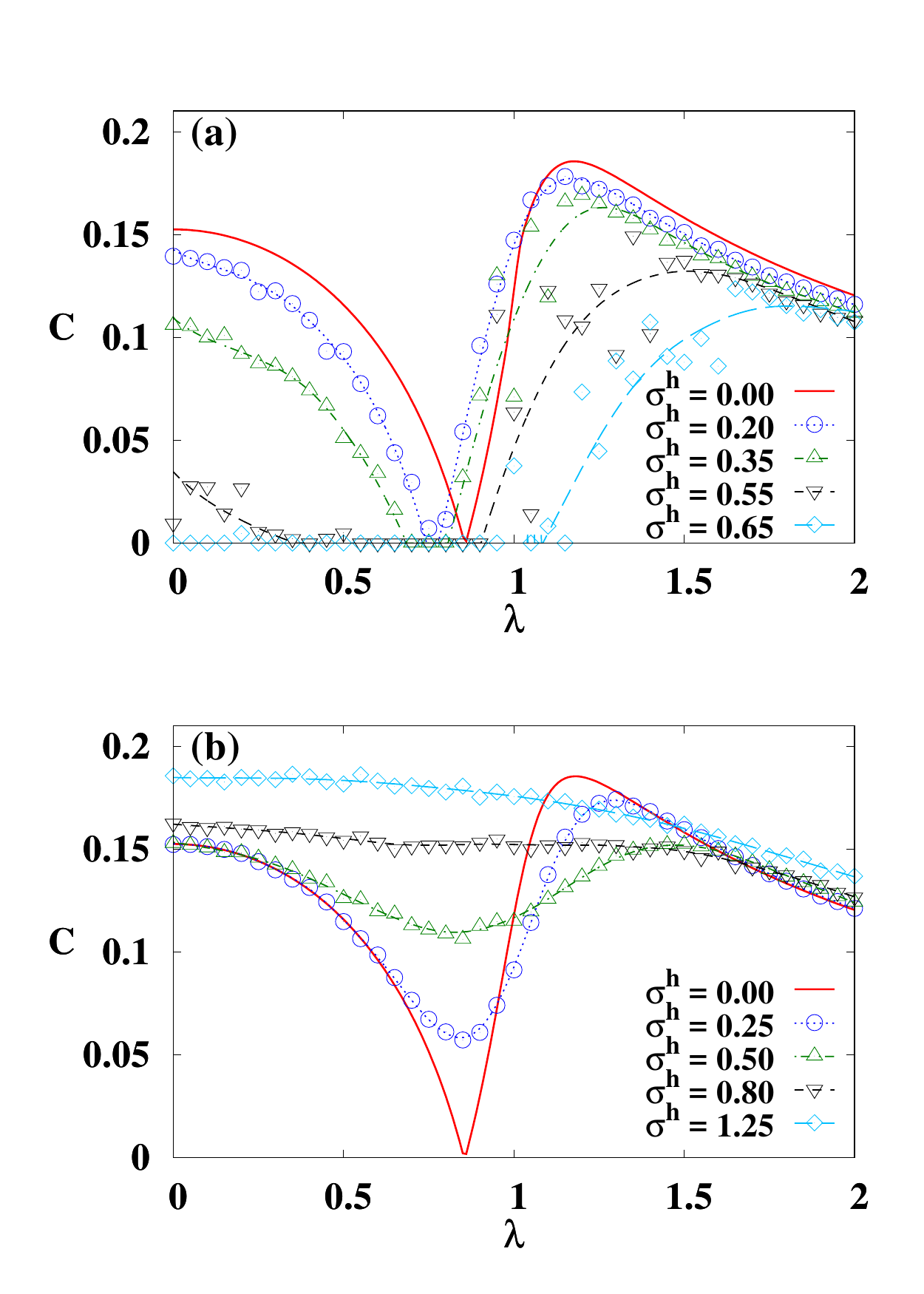}
\caption{(color online) Trends of entanglement in response to disorder in field. 
We plot the concurrence $C$, as a function of $\lambda= {\langle h\rangle}/\mathcal{J}$ for different disorder strengths, $\sigma^h$, in presence of (a)  annealed and (b)  quenched disorder in the transverse field. 
The concurrence is measured in ebits while $\lambda$ is dimensionless. Just like the previous figures, we choose $N=50$ and $\gamma = 0.5$ for Figs~\ref{fig_epsilonh}-\ref{fig_phase2h}. And, we choose $\beta {\cal J} = 20$ in these figures.
}
\label{fig_epsilonh}
\end{figure}

Let us now consider the quantum $XY$ spin chain with randomness present in the field term. In this case, nearest-neighbor exchange interactions are all equal in strength but $h_i$ are randomly chosen from i.i.d. Gaussian probability distributions with mean $\langle h \rangle$, and standard deviation $\tilde{\sigma}^h$. We consider the dimensionless quantity $\lambda = \langle h \rangle/{\cal J}$, and plot the concurrence with respect to the rescaled transverse field strength in Fig.~\ref{fig_epsilonh}. The corresponding dimensionless disorder strength is denoted as $\sigma^h = \tilde{\sigma}^h/{\cal J}$. 
Similar to Case 1, concurrence of the corresponding homogeneous system is computed at ${ h }/{\mathcal{J}}={\langle h \rangle}/{\mathcal{J}}=\lambda$. 
The homogeneous system remains entangled except at the factorization point, $\lambda = \lambda_f \equiv \sqrt{1-\gamma^2}$. 

\begin{figure}[t]
\vspace{0.8cm}
\includegraphics[angle=0,width=0.485\textwidth]{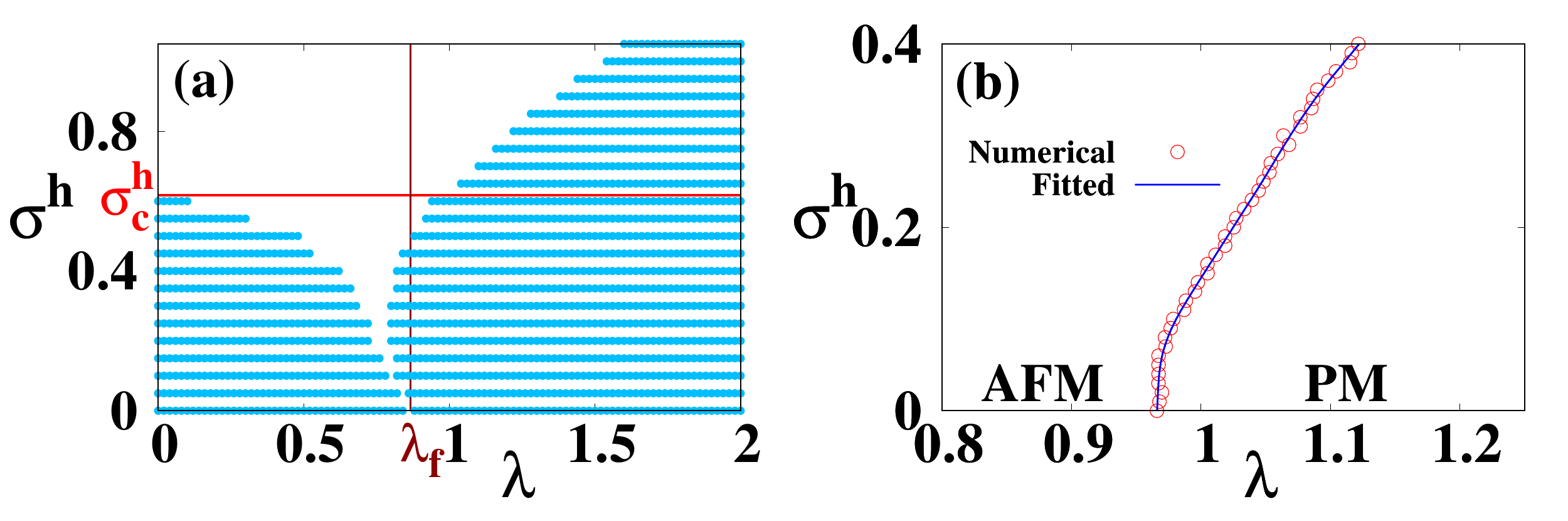}
\caption{An illustration of (a) entangled and separable phases in annealed disordered system, and (b) shifting of AFM-PM transition point in quenched disorder system. 
The (red) horizontal line in panel (a) represents $\sigma^h=\sigma^h_c$ and the (red) vertical line in the same panel represents $\lambda=\lambda_f$. The white region in panel (a) represents separable states, while the remaining regions represents entangled states. In panel (b), transition line is the (blue) solid line with (red) circles which is a fitted curve to the numerical data. The transition is characterized by a minimum of the derivative $\frac{\partial C}{\partial \lambda}$ as a function of $\lambda$ for a given $\sigma^h$. In the thermodynamic limit at zero temperature, this transition is at $\lambda=1$, which is slightly shifted in panel (b) due to finite size ($N=50$) and finite temperature $(\beta {\cal J} = 20)$. All axes in the panels represent dimensionless quantities. 
}
\label{fig_phase1h}
\end{figure}

The behavior of annealed  disordered concurrence with $\lambda$ for different values of disorder strength is shown in Fig. \ref{fig_epsilonh}(a).  
As we increase the strength of the annealed disorder,  the factorization point grows into a separable region which ultimately  swallows the entire AFM phase of the corresponding ordered system and also a significant portion of the PM phase, which is in contrast to the case when the disorder is present in the coupling strength. 
In the random field case, we identify a  critical disorder strength, at $\sigma^h=\sigma_c^h \approx 0.6$, below which there are two segments of entangled states on any line of constant $\sigma^h$. However, above the critical disorder strength, 
there is only a single sector of non-vanishing entanglement on any constant $\sigma^h$ line. 
We present the phase diagram of the separable and entangled phases in Fig. \ref{fig_phase1h}(a), which shows that in this case (just as in the case of annealed disordered interaction), high values of $\sigma^h$ destroys entanglement in both the magnetic phases of the corresponding ordered system.

It is interesting to mention here that the critical value of disorder strength for annealed disorder in the interaction
as well as the same in the field are both approximately the same, and \(\approx 0.6\).

We now consider the case of quenched disorder in the random field $XY$ model. Fig. \ref{fig_epsilonh}(b) shows the effect of quenched disorder on entanglement. Similar to the case of random interactions, here also  we find a drastic difference between the  entanglements for the two different kinds of disorder. However, the effect of quenched disorder is similar for both random interaction and random field cases. In both the scenarios, we find that with the increase of quenched disorder strength, entanglement over the entire parametric stretch of $\lambda$ tends to flatten and eventually gain similar value for all values of the control parameter, $\mu$ or $\lambda$. We observe that the introduction of a weak quenched disorder helps to grow entanglement even at the factorization point of the ordered system, resulting in the entire parameter space to support entangled states. This is unlike the annealed case, where the entire parameter space does not  become entangled for any finite amount of  disorder. We have also analyzed how the critical point of the corresponding homogeneous system shifts as we increase the disorder strength. The shifting point is again found from the extrema of the derivative of the concurrence. As plotted in Fig. \ref{fig_phase1h}(b), we find that minimum of the derivative of quenched disordered concurrence  shifts towards the PM region of the ordered system, as we 
increase the disorder strength, \(\sigma^h\). This is similar to our previous finding in Fig. \ref{fig_phase1}(b) for the system with quenched 
disordered random interaction, though the quantitative behaviors are different.

\begin{figure}[t]
\vspace{0.8cm}
\includegraphics[angle=0,width=0.485\textwidth]{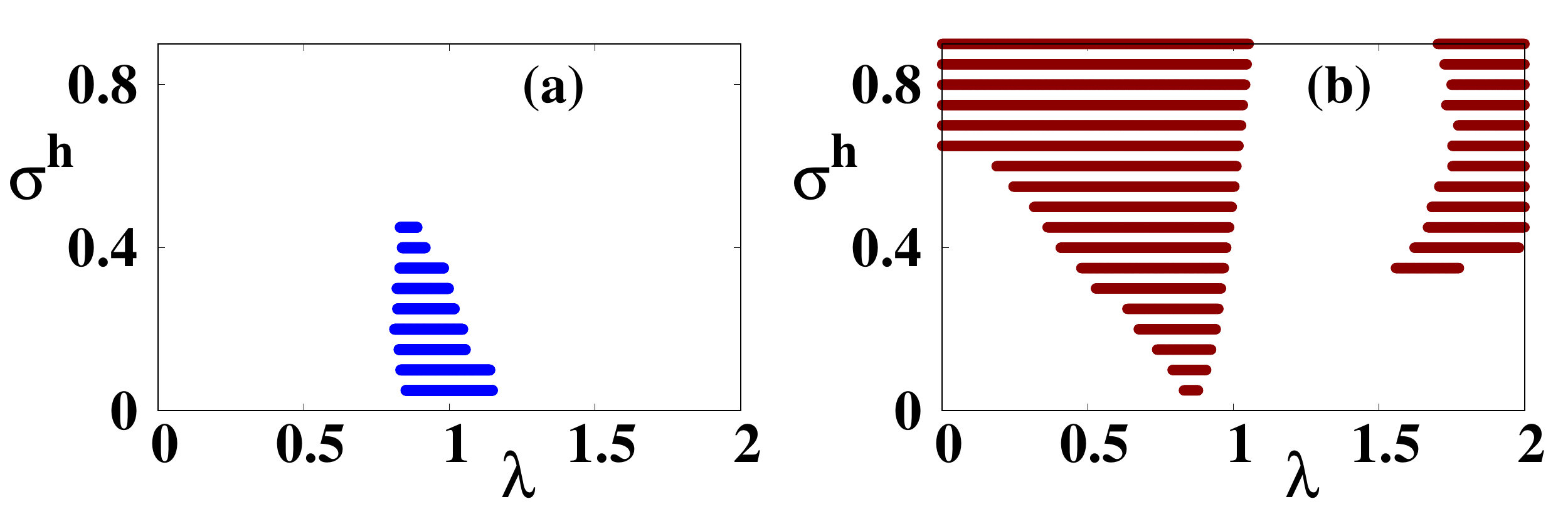}
\caption{An illustration of enhanced and normal phases in (a) annealed and (b) quenched disordered systems in the parameter space $(\lambda, \sigma^h)$. 
This is the parallel of Fig.~\ref{fig_OfD_phase} for the case of random fields instead of random couplings. The axes represent dimensionless quantities. 
}
\label{fig_phase2h}
\end{figure}

We complete our discussion by presenting the order-from-disorder analysis for the random field $XY$ model.  
In contrast to Case 1, here, the enhanced phase covers only a small parameter stretch in the case of annealed disorder as shown in Fig~\ref{fig_phase2h}(a). In particular, as we introduce the annealed disorder, an enhanced phase emerges only near to the critical point of the homogeneous system which quickly disappears with the increase of $\sigma^h$ 
However, as we increase quenched disorder strength, the enhanced phase grows and covers up a significant area in both the PM and AFM phases of the ordered systems (Fig.~\ref{fig_phase2h}(b)). With a high $\sigma^h$ value, the enhanced phase can be found in the entire AFM region and also in the PM region, except in the neighborhood of the critical point.  

\section{Conclusion}
\label{conclude}

This work aims to understand how the patterns of entanglement in a quantum many-body system at equilibrium are affected by insertion of disorder, and how they depend  on the  nature of disorder. 
We analyze two specific types of disorder, viz. annealed and quenched. We present  a general formalism for computing disorder averaged physical quantities for both kinds of disorder. While the formalism is valid for all temperatures, we focus on the physics at low temperatures. The models considered for the analyses are quantum  $XY$ spin chains with a transverse field, whose investigation have been carried out via the Jordan-Wigner and Bogoliubov transformations. Two particular cases are considered -- $XY$ chain with random interaction and $XY$ chain in random field. The disorders are chosen to be i.i.d. Gaussian random variables. Using concurrence as the quantifier of two-site entanglement, we studied its behavior for varying disorder strengths by tuning the standard deviations of the distribution functions.

Our analysis identifies the difference between the response of entanglement  due to quenched and annealed disorders.  We have shown that annealed disorder gives rise to  entangled and separable phases in the system. The factorization point of the corresponding ordered system grows into a separable phase in presence of annealed disorder. Moreover, for annealed disorder in the interaction, 
above a  certain critical value of disorder strength, any finite entanglement in the deep paramagnetic phase, of the corresponding ordered system, is washed out, and a finite value survives only in the deep antiferromagnetic phase. 
The same is valid for an annealed disorder in the field, but with the roles of the magnetic phases reversed.
On the contrary, entanglement is always non-vanishing in presence of quenched disorder. 

Presence of quenched as well as annealed disorder may exhibit the order-from-disorder phenomenon, by which we mean an enhancement of the bipartite entanglement with the introduction of disorder. There exist wide parameter regions, in both the magnetic phases of the corresponding ordered system, where disorder-induced enhancement occurs, for the quenched disordered systems. Such regions are relatively modest in area in the corresponding annealed systems. 

We also analyzed the effect on the quantum critical point in the corresponding ordered system due to the application of disorder. In particular, we found that with the increase of the quenched disorder strength, the  boundary between the magnetic phases of the ordered system, as quantified by the minimum of the derivative of entanglement, shifts towards the paramagnetic phase of the ordered system. 


Our work is relevant to currently available laboratory systems with engineered disorder having controllable disorder strength, particularly within an optical lattice set-up \cite{optical-lattice, Lewenstein_book}. Moreover, solid state systems with effective interaction strengths from a broad distribution, such as dilute magnetic semiconductors \cite{magnetic-semiconductor}, dislocation networks in solid $^4\text{He}$ \cite{solid-he}, provide platforms for accessing systems relevant to our work.

\acknowledgments
A.B. acknowledges the support of the Department of Science and Technology (DST), Govt. of India, through the award of an INSPIRE fellowship.
D.S. is supported in part by the `INFOSYS scholarship for senior students'.
D.R. acknowledges support from the EU Horizon 2020-FET QUIC 641122.


\begin{thebibliography}{100}

\bibitem{disoredr-review} K. Binder and A. P. Young, Rev. Mod. Phys. {\bf58}, 801 (1986); D. Belitz, T. R. Kirkpatrick, and T. Vojta, {\it ibid.} {\bf77}, 579 (2005); A. Das and B. K. Chakrabarti, {\it ibid.} {\bf80}, 1061 (2008); H. Alloul, J. Bobroff, M. Gabay, and P. J. Hirschfeld, {\it ibid.} {\bf81}, 45 (2009).

\bibitem{structural} D. J. Watts and S. H. Strogatz, Nature {\bf 393},440 (1998); A.-L. Barabasi and R. Albert, Science {\bf 286}, 509 (1999); A. Majdandzic, B. Podobnik, S. V. Buldyrev, D.Y. Kenett, S. Havlin and H. E. Stanley. Nature Physics {\bf 10}, 34 (2014).


\bibitem{loc1} P. W. Anderson, Basic Notions of Condensed Matter
Physics (Westview Press, Colorado, 1984); P. A. Lee and
T. V. Ramakrishnan, Rev. Mod. Phys. {\bf 57}, 287 (1985); R.
Zallen, The physics of amorphous solids (Wiley, New York,
1998).

\bibitem{loc2} P. W. Anderson, Phys. Rev. {\bf 109}, 1492 (1958); E. Abrahams,
P. W. Anderson, D. C. Licciardello, and T. V. Ramakrishnan,
Phys. Rev. Lett. {\bf 42}, 673 (1979).

\bibitem{loc3} A. Pal and D. A. Huse, Phys. Rev. B {\bf 82}, 174411 (2010);
M. Znidaric, T. Prosen, and P. Prelovsek, Phys. Rev. B
{\bf 77}, 064426 (2008); E. Canovi, D. Rossini, R. Fazio, G. E.
Santoro, and A. Silva, Phys. Rev. B {\bf 83}, 094431 (2011); J.
H. Bardarson, F. Pollmann, and J. E. Moore, Phys. Rev.
Lett. {\bf 109}, 017202 (2012); J. Eisert, M. Friesdorf, and C.
Gogolin, Nat. Phys. {\bf 11}, 124 (2015).

\bibitem{supercon} A. Auerbach, Interacting electrons and Quantum magnetism
(Springer, New York, 1994).

\bibitem{proclanation} D. Stauffer and A. Aharony, Introduction to Percolation Theory, 2nd edition (Taylor and Francis, Philadelphia, 1994).

\bibitem{phase1} D. Chowdhury, Spin Glasses and other Frustrated Systems (Wiley, New York, 1986); M. Mezard, G. Parisi, and M. A.
Virasoro, Spin Glass Theory and Beyond (World Scientific, Singapore, 1987).

\bibitem{phase2} S. Sachdev, Quantum Phase Transitions (Cambridge University
Press, Cambridge, 1999).

\bibitem{phase3} A. V. Goltsev, S. N. Dorogovtsev, and J. F. F. Mendes, Phys. Rev. E {\bf 67}, 026123 (2003); Z. Yao, K. P. C. da Costa, M. Kiselev, and N. Prokof'ev, Phys. Rev. Lett. {\bf 112}, 225301 (2014); J. P. {\'A}. {\'Z}{\~u}niga and N. Laflorencie, ibid. {\bf 111}, 160403 (2013).

\bibitem{optical-lattice} V. Ahufinger, L. S.-Palencia, A. Kantian, A. Sanpera, and M. Lewenstein, Phys. Rev. A {\bf 72}, 063616 (2005); M. Lewenstein, A. Sanpera, V. Ahufinger, B. Damski, A. Sen(De), and U. Sen, Adv. Phys. {\bf 56}, 243 (2007); L. Fallani, C. Fort, and M. Inguscio, Adv. At. Mol. Opt. Phys. {\bf 56}, 119 (2008); A. Aspect and M. Inguscio, Physics Today {\bf 62}, 30 (2009); L. S.-Palencia and M. Lewenstein, Nat. Phys. {\bf 6}, 87 (2010); G. Modugno, Rep. Prog. Phys. {\bf 73}, 102401 (2010); B. Shapiro, J. Phys. A {\bf 45}, 143001 (2012).

\bibitem{Lewenstein_book} M. Lewenstein, A. Sanpera, and V. Ahufinger, Ultracold atoms in Optical Lattices: simulating quantum many body physics (Oxford University Press, Oxford, 2012).

\bibitem{an-qun-0} L. F. Cugliandolo, Disordered systems, Lecture notes, (Cargese, 2011); S. G. Abaimov, Statistical Physics of Non-Thermal Phase Transitions, (Springer, 2015).

\bibitem{an-qun-1} C. De Dominicis and I. Giardina, Random Fields and Spin Glasses: A Field Theory Approach (Cambridge University Press, Cambridge, 2006).

\bibitem{an-qun-2} S. M.-Araghi and M. Sebtosheikh, Phys. Rev. E {\bf 92}, 022116 (2015).

\bibitem{an-qun-3} F. P. de A. Prado and G. M. Sch{\"u}tz, J. Stat. Phys. {\bf 142}, 984 (2011); A. N. M.-Kakkada, O. T. Valls, and C. Dasgupta, Phys. Rev. B {\bf 90}, 024202 (2014).

\bibitem{an-qun-4} R. Brout, Phys. Rev. {\bf 115}, 824 (1959).

\bibitem{Horodecki09} R. Horodecki, P. Horodecki, M. Horodecki, and K. Horodecki, Rev. Mod. Phys. {\bf 81}, 865 (2009).

\bibitem{ent-mb} M. Lewenstein, A. Sanpera, V. Ahufinger, B. Damski, A. Sen(De) and U. Sen, Adv. in Phys. {\bf 56}, 243 (2007); L. Amico, R. Fazio, A. Osterloh, and V. Vedral, Rev. Mod. Phys. {\bf 80}, 517 (2008).

\bibitem{dense-code} C. H. Bennett and S. J. Wiesner, Phys. Rev. Lett. {\bf 69}, 2881 (1992); K. Mattle, H. Weinfurter, P. G. Kwiat, and A. Zeilinger, Phys. Rev. Lett. {\bf 76}, 4656 (1996).

\bibitem{teleport} C. H. Bennett, G. Brassard, C. C{\'r}epeau, R. Jozsa, A. Peres, and W. K. Wootters, Phys. Rev. Lett. {\bf 70}, 1895 (1993); D. Bouwmeester, J. -W. Pan, K. Mattle, M. Eibl, H. Weinfurter, and A. Zeilinger, Nature {\bf 390}, 575 (1997).

\bibitem{computation} R. Raussendorf and H. J. Briegel, Phys. Rev. Lett. {\bf 86}, 5188 (2001); F. Meier, J. Levy, and D. Loss, ibid.{\bf 90}, 047901 (2003); R. Raussendorf, D. E. Browne, and H. J. Briegel, Phys. Rev. A {\bf 68}, 022312 (2003); M. A. Nielsen, Phys. Rev. Lett. {\bf 93}, 040503 (2004); P. Walther, K. J. Resch, T. Rudolph, E. Schenck, H. Weinfurter, V. Vedral,
M. Aspelmeyer, and A. Zeilinger, Nature {\bf 434}, 169 (2005); M. A. Nielsen, Rep. Math. Phys. {\bf 57}, 147 (2006); H. J. Briegel, D. E. Browne, W. D{\"u}r, R. Raussendorf, and M. V. den Nest, Nat. Phys. {\bf 5}, 19 (2009).

\bibitem {quantum-key}A. K. Ekert, Phys. Rev. Lett. {\bf 67}, 661 (1991); N. Gisin, G. Ribordy, W. Tittel, and H. Zbinden, Rev. Mod. Phys. {\bf 74}, 145 (2002).



\bibitem{ent-phase} T. J. Osborne and M.A. Nielsen, Quantum Inf. Proc. {\bf 1}, 45 (2002); T. J. Osborne and M. A. Nielsen, Phys. Rev. A {\bf 66}, 032110 (2002); A. Osterloh, L. Amico, G. Falci, and R. Fazio, Nature {\bf 416}, 608 (2002).



\bibitem{ent-supercon}  S. Oh and J. Kim, Phys. Rev. B {\bf 71}, 144523 (2005); P. Sacramento, P. Nogueira, V. Vieira, and V. Dugaev, Phys. Rev. B {\bf 76} 184517 (2007); T. P. Oliveira and P. D. Sacramento, Phys. Rev. B {\bf 89}, 094512 (2014).

\bibitem{Dziarmaga06} J. Dziarmaga, Phys. Rev. B {\bf 74}, 064416 (2006); T. Caneva, R. Fazio and G. E. Santoro, Phys. Rev. B {\bf 76}, 144427 (2007).

\bibitem{wehr} M. Aizenman and J. Wehr, Phys. Rev. Lett. {\bf 62}, 2503 (1989); M. Aizenman and J. Wehr, Comm. Math. Phys.
{\bf 130}, 489 (1990); J. Wehr, A. Niederberger, L. S.-Palencia, and M. Lewenstein, Phys. Rev. B {\bf 74}, 224448 (2006).

\bibitem{disorder-spin} D. Burgarth and S. Bose, New J. Phys. {\bf 7}, 135 (2005); C. K. Burrell and T. J. Osborne, Phys. Rev. Lett. {\bf 99}, 167201 (2007); C. K. Burrell, J. Eisert and T. J. Osborne, Phys Rev A {\bf 80} 052319 (2009); J. Allcock and N. Linden Phys. Rev. Lett. {\bf 102}, 110501 (2009); D. Petrosyan, G. M. Nikolopoulos and P. Lambropoulos,Phys. Rev. A {\bf 81}, 042307 (2010); X. Wang, A. Bayat, S. G. Schirmer and S. Bose, Phys. Rev. A {\bf 81}, 032312 (2010).

\bibitem{qn-entanglement} G. De Chiara, S. Montangero, P. Calabrese and R. Fazio, J. Stat. Mech. 03001  (2006); M. Fujinaga and N. Hatano, J. Phys. Soc. Jpn. {\bf 76}, 094001 (2007); D. Binosi, G. De Chiara, S. Montangero, and A. Recati, Phys. Rev. B {\bf 76}, 140405(R) (2007).

\bibitem{Sadhukhan} D. Sadhukhan, R. Prabhu, A. Sen(De), U. Sen, Phys. Rev. E {\bf 93}, 032115 (2016).

\bibitem{group} R. Prabhu, S. Pradhan, A. Sen(De), and U. Sen, Phys. Rev. A {\bf 84}, 042334 (2011); U. Mishra, D. Rakshit, R. Prabhu, A. Sen(De), and U. Sen, New J. Phys. {\bf 18}, 083044 (2016).

\bibitem{ent-lgt} D. Sadhukhan, S. Singha Roy, D. Rakshit, R. Prabhu, A. Sen(De), and U. Sen Phys. Rev. E {\bf 93}, 012131 (2016).

\bibitem{no-go} D. Sadhukhan, S. Singha Roy, D. Rakshit, A. Sen(De), and U. Sen, New J. Phys. {\bf 17}, 043013 (2015).

\bibitem{thorpe} M. F. Thorpe and D. Beeman, Phys. Rev. B {\bf 14}, 188 (1976); M. F. Thorpe and D. Beeman, Phys. Rev. B {\bf 14}, 188 (1976); M. F. Thorpe, J. Phys. C: Solid State Phys., {\bf 11}, 2983 (1978).

\bibitem{dasgupta} A. N. Malmi-Kakkada, O. T. Valls, and C. Dasgupta, Phys. Rev. B {\bf 90}, 024202 (2014).

\bibitem{bera2} A. Bera, D. Rakshit, A. Sen(De), U. Sen,  Phys. Rev. B {\bf 95}, 224441 (2017).

\bibitem{hide1} J. Hide, J. Phys. A: Math. Theor. {\bf 45}, 115302 (2012).

\bibitem{hide2} J. Hide, W. Son, and V. Vedral, Phys. Rev. Lett. {\bf 102}, 100503 (2009).


\bibitem{lieb61} E. Lieb, T. Schultz, and D. Mattis, Ann. Phys. {\bf 16}, 407 (1961).

\bibitem{Barouch70} E. Barouch, B. McCoy, and M. Dresden, Phys. Rev. A {\bf 2}, 1075 (1970); E. Barouch and B. McCoy, Phys. Rev. A {\bf 3}, 786 (1971).

\bibitem{factor1} J. Kurmann, H. Thomas, and G. Muller, Physica A 112, 235 (1982); G. M{\"u}ller, R. E. Shrock, Phys. Rev. B {\bf 32},
5845 (1985).

\bibitem{factor2} J. Eakins and G. Jaroszkiewicz, J. Phys. A: Math. Gen. {\bf 36}, 517 (2003); T. Roscilde, P. Verrucchi, A. Fubini, S. Haas, and V. Tognetti, Phys. Rev. Lett. {\bf 93}, 167203 (2004), {\it ibid.} {\bf 94}, 147208 (2005); S. Dusuel and J. Vidal, Phys. Rev. B {\bf 71}, 224420 (2005); L. Amico, F. Baroni, A. Fubini, D. Patan, V. Tognetti, and Paola Verrucchi,
Phys. Rev. A {\bf 74}, 022322 (2006); F. Baroni, J. Phys. A {\bf 40}, 9845 (2007); F. Baroni, A. Fubini, V. Tognetti and
P. Verrucchi, ibid. {\bf 40}, 9845 (2007); S. M. Giampaolo, G. Adesso, and F. Illuminati. Phys. Rev. Lett. {\bf 100},
197201 (2008); B. Cakmak, G. Karpat and F. F. Fanchini, arXiv:1502.02306 (2015).

\bibitem{od-dis} A. Aharony, Phys. Rev. B {\bf 18}, 3328 (1978); J. Villain,
R. Bidaux, J.-P. Carton, and R. Conte, J. Physique {\bf 41},
1263 (1980); B. J. Minchau and R. A. Pelcovits, Phys.
Rev. B {\bf 32}, 3081 (1985); C. L. Henley, Phys. Rev. Lett.
{\bf 62}, 2056 (1989); A. Moreo, E. Dagotto, T. Jolicoeur, and
J. Riera, Phys. Rev. B {\bf 42}, 6283 (1990); D. E. Feldman,
J. Phys. A {\bf 31}, L177 (1998); G. E. Volovik, JETP Lett.
{\bf 84}, 455 (2006); D. A. Abanin, P. A. Lee, and L. S. Levitov,
Phys. Rev. Lett. {\bf 98}, 156801 (2007); L. Adamska,
M. B. Silva Neto, and C. Morais Smith, Phys. Rev. B
{\bf 75}, 134507 (2007); A. Niederberger, T. Schulte, J. Wehr,
M. Lewenstein, L. Sanchez-Palencia, and K. Sacha, Phys.
Rev. Lett. {\bf 100}, 030403 (2008); A. Niederberger, J.Wehr,
M. Lewenstein, and K. Sacha, Europhys. Lett. {\bf 86}, 26004
(2009); A. Niederberger, M. M. Rams, J. Dziarmaga, F.
M. Cucchietti, J. Wehr, and M. Lewenstein, Phys. Rev. A
{\bf 82}, 013630 (2010); D. I. Tsomokos, T. J. Osborne, and C.
Castelnovo, Phys. Rev. B {\bf 83}, 075124 (2011); M. S. Foster,
H.-Y. Xie, and Y.-Z. Chou, ibid. {\bf 89}, 155140 (2014);
P. V. Mart\'in, J. A. Bonachela, and M. A. Munoz, Phys.
Rev. E {\bf 89}, 012145 (2014).

\bibitem{bera} A. Bera, D. Rakshit, M. Lewenstein, A. Sen(De), U. Sen, J. Wehr, Phys. Rev. B {\bf 90}, 174408 (2014); A. Bera, D.
Rakshit, M. Lewenstein, A. Sen(De), U. Sen, J. Wehr, Phys. Rev. B {\bf 94}, 014421 (2016).

\bibitem{hill} S. Hill and W. K. Wootters, Phys. Rev. Lett. {\bf 78}, 5022 (1997); W. K. Wootters, ibid. {\bf 80}, 2245 (1998).



\bibitem{magnetic-semiconductor} V. M. Galitski, A. Kaminski, and S. Das Sarma, Phys. Rev. Lett. {\bf 92}, 177203 (2004); A. Kaminski and S. Das Sarma, Phys. Rev. Lett. {\bf 88}, 247202 (2002).

\bibitem{solid-he} L. Pollet, M. Boninsegni, A. B. Kuklov, N. V. Prokof'ev, B. V. Svistunov and M. Troyer, Phys. Rev. Lett. {\bf 98}, 135301 (2007); M. Boninsegni, A. B. Kuklov, L. Pollet, N. V. Prokof'ev, B. V. Svistunov and M. Troyer, Phys. Rev. Lett. {\bf 99}, 034301 (2007); S. I. Shevchenko, Fiz. Nauk. Temp. {\bf 14} (1988) [Sov. J. Low Temp. Phys {\bf 14}, 553 (1988)]; A. T. Dorsey, P. M. Goldbart, and J. Toner, Phys. Rev. Lett. {\bf 96}, 055301 (2006); J. Toner, Phys. Rev. Lett. {\bf 100}, 35302 (2008); C. Dasgupta and O. T. Valls, Phys. Rev. B {\bf 82}, 024523 (2010).

\end{thebibliography}
\end{document}